\documentclass[a4paper,fleqn,useAMS,usenatbib]{mnras}

\newcommand{\be}{\begin{equation}}
\newcommand{\ee}{\end{equation}}
\newcommand{\ba}{\begin{eqnarray}}
\newcommand{\ea}{\end{eqnarray}}

\def\simless{\mathbin{\lower 2.5pt\hbox
   {$\rlap{\raise 4.5pt\hbox{$\char'074$}}\mathchar"7218$}}}
\def\simgrt{\mathbin{\lower 2.5pt\hbox
   {$\rlap{\raise 4.5pt\hbox{$\char'076$}}\mathchar"7218$}}}
\include{epsf}

\title[UHS Definition and $J$-band Data Release]
{The UKIRT Hemisphere Survey: Definition and $J$-band 
Data Release}
\author[S. Dye et al.]
{S. Dye,$^1$\thanks{E-mail: simon.dye@nottingham.ac.uk}
A. Lawrence$^2$, M. A. Read$^2$,
X. Fan$^3$, T. Kerr$^4$, W. Varricatt$^4$, \newauthor
K. E. Furnell$^5$,
A. Edge$^6$, M. Irwin$^7$, N. Hambly$^2$,
P. Lucas$^8$, O. Almaini$^1$, \newauthor
K. Chambers$^9$, R. Green$^3$, 
P. Hewett$^7$, M. C. Liu$^9$,
I. McGreer$^3$, W. Best$^9$, \newauthor
Z. Zhang$^9$, E. Sutorius$^2$, 
D. Froebrich$^{10}$, E. Magnier$^9$, G. Hasinger$^9$, \newauthor
S. M. Lederer$^{11}$, M. Bold$^{12}$, J. A. Tedds$^{13}$
\vspace{4mm}\\
$^1$School of Physics and Astronomy, Nottingham University,
University Park, Nottingham, NG7 2RD, UK\\
$^2$Institute for Astronomy, Blackford Hill, Edinburgh, EH9 3HJ, UK\\
$^3$Department of Astronomy/Steward Observatory, 933 North Cherry 
Avenue, Tucson, AZ 85721-0065, USA\\
$^4$UKIRT, 660 N. A'oh\={o}k\={u} Place, University Park,
Hilo, Hawaii 96720, USA\\
$^5$Astrophysics Research Institute, LJMU, IC2, Liverpool 
Science Park, 146 Brownlow Hill, Liverpool, L3 5RF, UK\\
$^6$Department of Physics, University of Durham, South Road, 
Durham, DH1 3LE, UK\\
$^7$Institute of Astronomy, Madingley Road, Cambridge, CB3 0HA, UK\\
$^8$School of Physics, Astronomy and Mathematics, University 
of Hertfordshire Hatfield Hertfordshire, AL10 9AB, UK\\
$^9$Institute for Astronomy, University of Hawaii,
2680 Woodlawn Drive, Honolulu, Hi 96822, USA\\
$^{10}$Centre for Astrophysics and Planetary Science, University of Kent, 
Canterbury, CT2 7NH, UK\\
$^{11}$NASA JSC, Mail Code: XI, 2101 NASA Parkway, Houston, TX 77058, USA\\
$^{12}$Lockheed Martin Space Systems Company, 1111 Lockheed Martin Way,
Sunnyvale, CA 94089, USA\\
$^{13}$Department of Physics and Astronomy, University of Leicester, 
Leicester, LE1 7RH, UK
}

\begin{document}

\date{}

\pagerange{\pageref{firstpage}--\pageref{lastpage}} 
\pubyear{2017}

\maketitle

\label{firstpage}

\begin{abstract}
This paper defines the UK Infra-Red Telescope (UKIRT) Hemisphere
Survey (UHS) and release of the remaining $\sim$12,700 deg$^2$ of
$J$-band survey data products. The UHS will provide continuous $J$ and
$K$ band coverage in the northern hemisphere from a declination of
0$^\circ$ to 60$^\circ$ by combining the existing Large Area Survey,
Galactic Plane Survey and Galactic Clusters Survey conducted under the
UKIRT Infra-red Deep Sky Survey (UKIDSS) programme with this new
additional area not covered by UKIDSS. The released data includes $J$
band imaging and source catalogues over the new area, which, together
with UKIDSS, completes the $J$-band UHS coverage over the full
$\sim$17,900 deg$^2$ area. 98 per cent of the data in this release
have passed quality control criteria, the remaining 2\,per cent being
scheduled for re-observation. The median 5$\sigma$ point source
sensitivity of the released data is 19.6\,mag (Vega). The median full
width at half-maximum of the point spread function across the dataset
is 0.75\,arcsec. In this paper, we outline the survey management, data
acquisition, processing and calibration, quality control and archiving
as well as summarising the characteristics of the released data
products. The data are initially available to a limited consortium
with a world-wide release scheduled for August 2018.

\end{abstract}


\section{Introduction}

Large area surveys play a fundamental role in astronomy. They are an
efficient use of resources; in Euclidean space, a shallower and wider
area survey explores a larger volume than a deeper and narrower one
for the same total exposure time. They represent a basic point of
reference, providing the statistical samples needed to address many
intrinsically large angular area problems in astronomy such as
studying the structure of the Milky Way and investigating the global
structure of the Universe.  Large area surveys are also a source of
unexpected and serendipitous discoveries of rare and new classes of
objects, such as high redshift quasars \citep[see][]{wang17,yang17},
ultra-cool brown dwarfs and free-floating planets \citep[see, for
  example][]{liu13,marocco14}. Full-sky or near full-sky survey
coverage brings about the capability of retrospective identification
of sources, for example, following a rare event or to find
counterparts to sources detected at longer wavelengths where
resolution is poor.

This paper defines the United Kingdom Infra-Red Telescope (UKIRT)
Hemisphere Survey (UHS) and the release of data which complete the
survey's $J$-band coverage. The survey is the product of an
evolving collaboration between several international partners. During
the period of acquisition of data in this release, these partners have
been the UK Science and Technology Facilities Council (STFC), The
University of Hawaii, The University of Arizona, Lockheed Martin and
the National Aeronautics and Space Administration (NASA).

Survey data were released via the Wide Field Camera
\citep[WFCAM;][]{casali07} Science
Archive \footnote{http://wsa.roe.ac.uk} \citep[WSA; see][]{hambly08}
on August 1st 2017. This release is initially to
UK, University of Hawaii and University of Arizona astronomers.  A
world-wide release will occur in one year.  The release incorporates
all data observed from 19th May 2012 up to and including January 31st
2017. This constitutes $\sim$ 12,700\,deg$^2$ of $J$
band imaging and associated source catalogue products reaching a
median depth of 19.6\,mag (Vega) and containing approximately 500
million detected sources.

The aim of the UHS is to provide continuous coverage in the northern
hemisphere in the $J$ and $K$-band over the declination range of
$0^\circ \leq \delta \leq 60^\circ$, this upper limit being set by
UKIRT's mechanical constraints.  This goal is achieved by surveying
the area within this declination range that is not already covered by
the three large area surveys carried out within the UKIRT Infra-red
Deep Sky Survey \citep[UKIDSS;][]{lawrence07} programme. These three
surveys are the Large Area Survey (LAS) covering an area of $\sim
3700$\,deg$^2$, the Galactic Plane Survey \citep[GPS;][]{lucas08}
which covers $\sim 2100$\,deg$^2$ along the northern and equatorial
Galactic plane and the Galactic Clusters Survey (GCS) with an area of
$\sim 400$\,deg$^2$. Together, the LAS, GPS and GCS provide
$\sim$5,200\,deg$^2$ of the full UHS area\footnote{Note that the LAS
  extends to declinations $<0^\circ$ and that there is overlap in
  coverage between the LAS, GCS and GPS.}.

The coverage of the UHS (see section \ref{sec_geom} and Figure
\ref{maps}) complements that of the Visible and Infrared
Survey Telescope for Astronomy (VISTA) Hemisphere Survey
\citep[VHS;][]{mcmahon13} which covers $\sim 18,000$\,deg$^2$ in the
southern hemisphere in $J$ and $K_s$ to similar depths as the UHS. The
combination of UHS and VHS comprises an almost complete near-infrared
sky survey $\sim 4$\,mag deeper than the Two-Micron All-Sky Survey
\citep[2MASS;][]{skrutskie06}, thus giving rise to a long-term legacy
database. In addition, the Wide-field Infrared Survey (WISE) all-sky
catalogue \citep{wright10} which extends to mid-infrared wavelengths
and the optical coverage provided by both the Panoramic Survey
Telescope \& Rapid Response System (Pan-STARRS) survey
\citep{chambers16} and the Sloan Digital Sky Survey \citep[SDSS;
  see][for the 13th data release]{albareti17} realise an extremely
powerful set of surveys covering two decades in wavelength and with
well-matched sensitivities.


The layout of this paper is as follows. In Section \ref{sec_design},
we outline the survey geometry, observing strategy, pipeline
processing, calibration and how the survey products are archived and
accessed at the WSA.  Section \ref{sec_products} details the data
products made available along with supplementary information specific
to this release. Information on data quality control (QC) procedures
and characteristics of the data are given in section
\ref{sec_data_props}. Finally, we summarise in section
\ref{sec_summary}.  All magnitudes quoted in this paper are Vega
magnitudes in the system described by \citet{hewett06}.

\section{Survey Design}
\label{sec_design}

UKIRT is an infrared-specific telescope with a primary mirror of
diameter 3.8\,m. Together, UKIRT and WFCAM offer an impressive
\'{e}tendue of $\sim 2.4$\,m$^2$deg$^2$, defined as the product of the
telescope collecting area and the solid angle of the camera's field of
view (FOV) . This, combined with low observing overheads due to the
high efficiency of WFCAM and a fast tip-tilt secondary mirror on UKIRT
which allows dither offsets within 10\,arcsec to be carried out
without slewing the telescope, makes a highly optimised infrared
survey facility.

\begin{figure}
\epsfxsize=85mm
{\hfill
\epsfbox{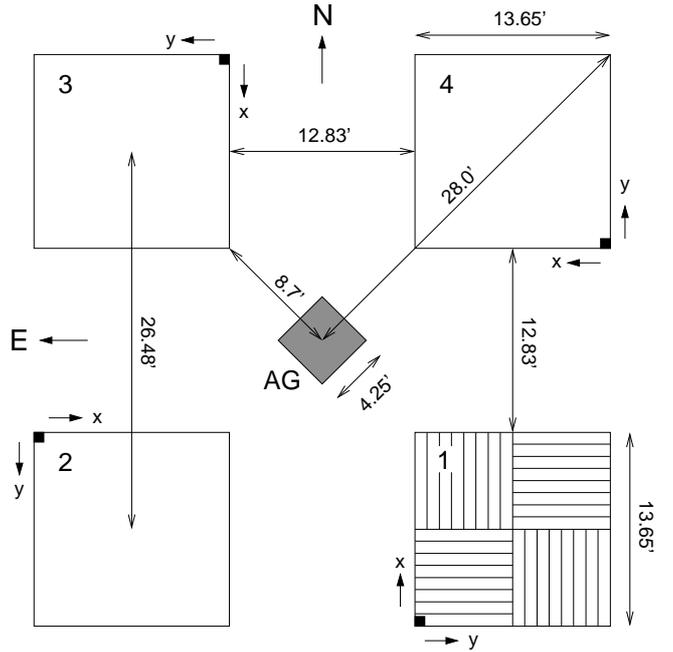}
\hfill}
\caption{The WFCAM focal plane showing the four detectors and the
  autoguider (AG). Numbering of detectors is the same as that used in
  the WSA and for extension numbering in the multi-extension FITS
  image files. Detectors are divided into four quadrants, each
  quadrant having 8 channels as illustrated in detector 1. Note the
  convention used for the image pixel co-ordinate system in each
  detector.  In this figure taken from \citet{dye06}, north is upwards
  and east is to the left. }
\label{wfcam_focal_plane}
\end{figure}

In this paper, we adopt the nomenclature used by UKIDSS. Accordingly,
an {\it Exposure} is a single 10\,s integration which gives rise to a
single data file comprising the four separate detector images.  The
stack of four 10\,s Exposures results in a {\it Stack} frame.  The
centres of the detector images in an Exposure or Stack frame are
offset from the `base-position' at the frame centre by 18.72\,arcmin
in NE, NW, SE and SW directions.  We use the term {\it Tile} to
designate the contiguous square area formed by positioning four Stacks
in a square $2 \times 2$ grid of spacing 13.24\,arcmin (see below). We
maintain capitalisation hereafter to indicate instances of these
specific definitions.

Figure \ref{wfcam_focal_plane} \citep[taken from][for
  completeness]{dye06}, shows the layout of the focal plane of
WFCAM. An optical charge coupled device acting as the auto-guider lies
at the centre of the focal plane. The four square $2048 \times 2048$
pixel detectors are spaced at 94 per cent of the 13.65\,arcmin angular
width they subtend on the sky.  This gives an areal coverage of
0.21\,deg$^2$ and 0.79\,deg$^2$ per Exposure and per Tile
respectively. Readers are referred to \citet{casali07} for further
technical details regarding WFCAM.

\subsection{Geometry and tiling}
\label{sec_geom}

\begin{figure*}
\epsfxsize=180mm
{\hfill
\epsfbox{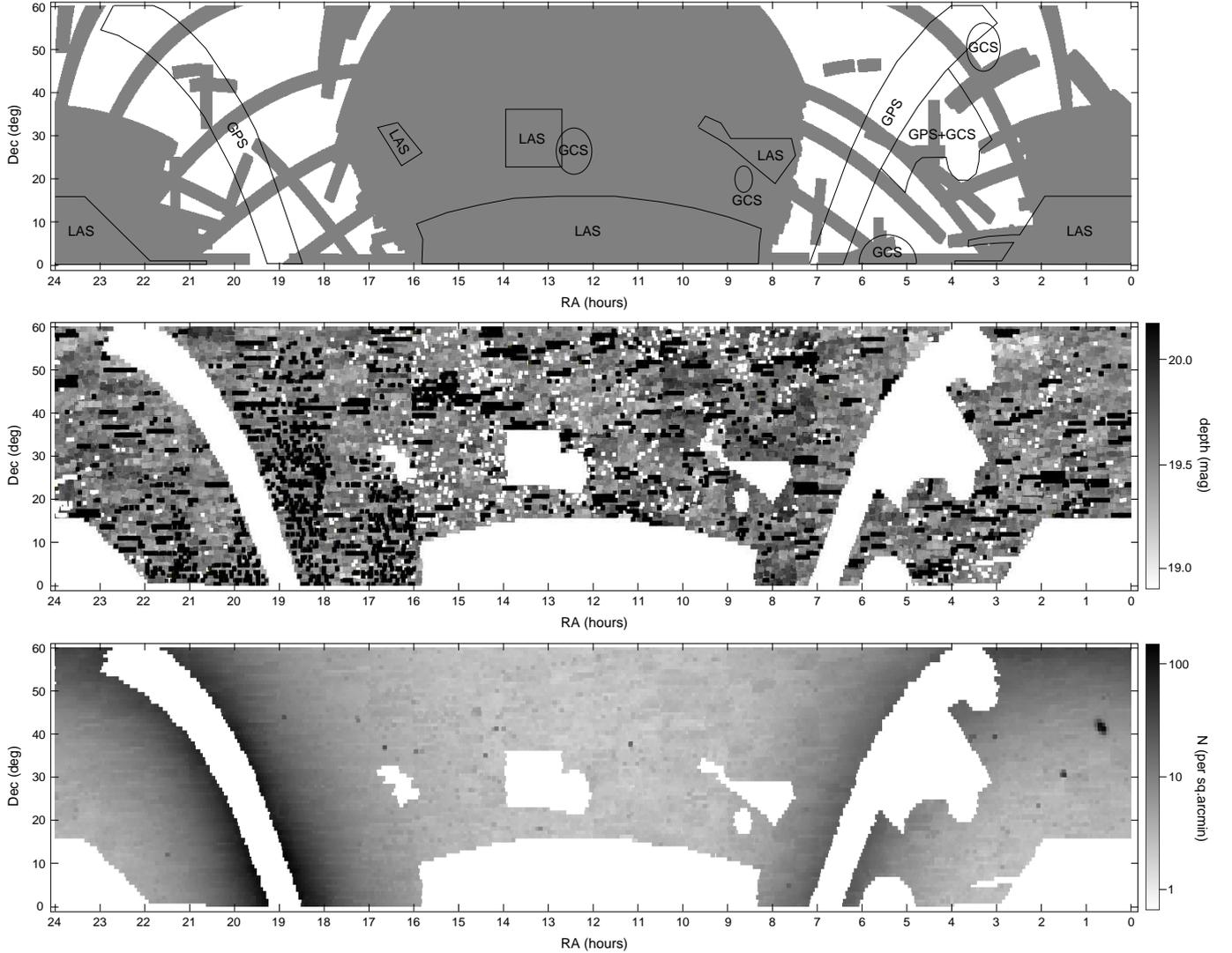}
\hfill}
\caption{{\em Top:} Survey Geometry. The UHS covers the full area
  plotted, with the 'rUHS' area (constituting the present release)
  being defined as that area not covered by the UKIDSS LAS, GPS or
  GCS regions as indicated. Coverage of imaging in SDSS DR13
  \citep{albareti17} is also plotted in grey for comparison. {\em
    Middle:} A map of 5$\sigma$ $J$-band point source sensitivity in
  rUHS. Black squares indicate multiframes which have been imaged at
  least twice and white squares indicate multiframes not included in
  the standard release products (see Section \ref{sec_dr1_specifics}
  for more details).  {\em Bottom:} Number density of sources detected
  in all multiframes in rUHS, including the extended release
  products.}
\label{maps}
\end{figure*}

The UHS geometry is plotted in Figure \ref{maps} which shows the
existing area covered by the LAS, GPS and GCS. Hereafter, we refer to
the area not covered by these UKIDSS surveys as `rUHS', the `r'
designating the `remaining' area.

The rUHS area was tiled using a bespoke piece of software called the
Survey Definition Tool (SDT) originally developed for UKIDSS. The SDT
forms a contiguous area by overlapping Tiles in right
ascension and declination with a nominal overlap of $\sim 0.5$\,arcmin
in both directions,
set to match the overlap between Exposures within a Tile. Tile
placement is also constrained by the availability of good guide stars
which must lie within the area of sky subtended by the autoguider's
FOV and be within user-defined magnitude limits. For the rUHS, these
magnitude limits are $11.0<B_J<16.5$ in photographic $B_J$ band which has
an effective wavelength of $\sim 470$\,nm. If a suitable guide star is
not identified for all four Exposures within a Tile, then the SDT
attempts to back-track in RA to find one. This results in the full
Tile overlapping its neighbouring Tile by $>0.5$\,arcmin.
Backtracking can continue until the Tile overlaps its neighbouring
tile by 75\,per cent.  If this backtracking still does not result in
finding a guide star, the SDT attempts to place two tiles back at the
nominal RA but one at a higher declination and one at a lower
declination than the nominal declination.  These are given the
designation 'north' and 'south' (stored as the {\tt object} attribute
of table Multiframe in the WSA - see section \ref{sec_data_access}).  If
guide stars are still not found, then these two Tiles are backtracked
until one is.

This scheme proves successful for the rUHS geometry with all Tiles
being assigned guide stars without leaving any areas uncovered. 
However, the SDT selects its guide stars from version 2.2 of the
Hubble Space Telescope's guide star catalogue \citep{lasker08} where a
small fraction of sources are actually compact galaxies. These compact
galaxies prove problematic for the auto guider which cannot reliably
locate their centroid. To determine occurrences of compact galaxies,
the SDSS was used where possible and when not, the WISE all-sky survey
catalogue and the 2MASS catalogue were used. It was found that the
incorrectly selected compact galaxies tend to have red $J-K$ 2MASS
colours and lie red-ward of the stellar locus in the WISE colour
$m(3.4\,\mu{\rm m})-m(4.6\,\mu{\rm m})$.  Appropriate colour
thresholds therefore identified the majority of contaminating compact
galaxies which were replaced manually, occasionally requiring Exposure
frames to be offset from their nominal position within problematic
Tiles.  Despite these efforts, a small fraction of compact galaxies
have remained within the selection of guide stars and this is evident
from the measured increase in the fraction of images that have failed
due to trailing with increasing galactic latitude (see Section
\ref{sec_data_quality}).

\begin{figure}
\epsfxsize=80mm
{\hfill
\epsfbox{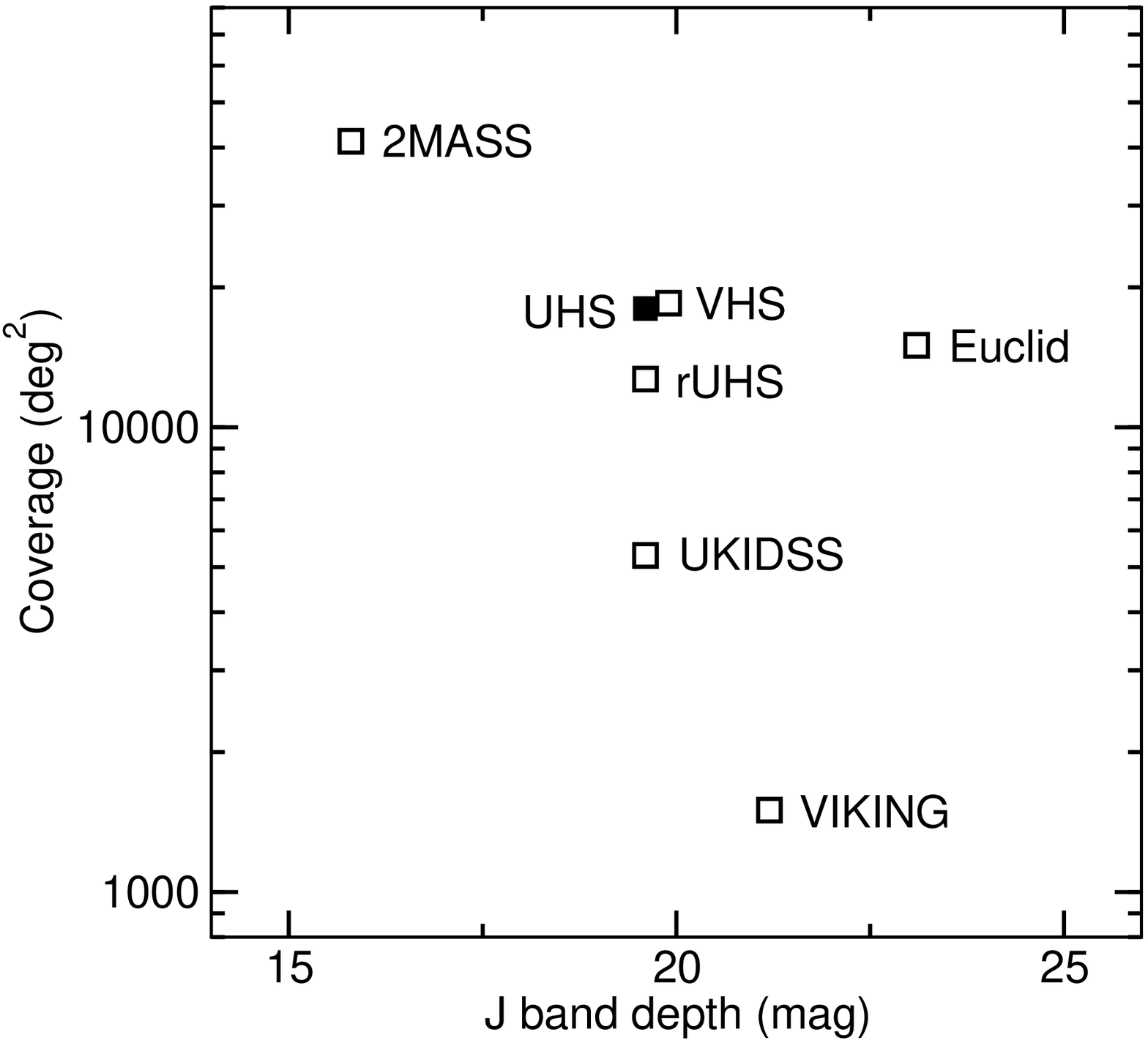}
\hfill}
\caption{Comparison of 5$\sigma$ point source sensitivities and
  coverage of completed, ongoing and forthcoming large area (i.e.
  $>1000$\,deg$^2$) near-IR surveys. Statistics are shown for 2MASS,
  UKIDSS (LAS+GPS+GCS), VHS, the VISTA Kilo-degree Infrared Galaxy
  survey \citep[VIKING;][]{edge13} and the Euclid wide area survey
  \citep[survey parameters taken from][]{racca16} as well as the UHS
  and rUHS. The UHS has almost the same areal coverage as the VHS, but
  in the opposite hemisphere.}
\label{survey_comparison}
\end{figure}

Tiling of the rUHS area was carried out by initially tiling the
complete northern hemisphere up to a declination of 60$^\circ$ and
then removing Tiles coincident with those of the LAS, GPS and GCS as
defined in the 13th UKIDSS data release.  This was carried out in such
a way as not to leave any gaps at UKIDSS boundaries. The rUHS area
requires $\sim$16,500 Tiles for complete coverage, which, allowing for
overlaps between Tiles, constitutes an area of
$\sim$12,700\,deg$^2$. For scheduling purposes, the rUHS area is split
into 24 projects, each project spanning one hour in right
ascension. Project names appear as the attribute {\tt project} in the
{\tt Multiframe} table in the WSA.

\subsection{Observing strategy}

Observations are divided into `minimum schedulable blocks'
(MSBs). Each rUHS MSB has an execution time of approximately 30
minutes, designed to be short enough to maintain flexibility of
scheduling with other projects at UKIRT but long enough to obtain a
sufficient number of observations for generation of the sky
subtraction frame (see section \ref{sec_pipeline}). This ensures that
MSBs are self contained.  MSBs typically complete eight Tiles
totalling an on-sky integration time of 1280\,s and therefore an
overall overhead fraction of 28 per cent due to telescope slewing
and detector read-out and flushing \citep[see][]{lawrence07}.

MSBs are only executed if the site quality criteria are met. For the
rUHS, these are a maximum $J$ band sky brightness of
15.5\,mag.arcsec$^{-2}$, a maximum full width at half-maximum (FWHM)
of the point spread function\footnote{Due to optical aberrations and
  imperfect alignment between components within WFCAM, the delivered
  image PSF FWHM of 1.2\,arcsec actually corresponds to an intrinsic
  seeing of $\sim 1.15$\,arcsec.} (PSF) of 1.2\,arcsec and the
requirement that conditions are photometric.  The quality of
observations can fall outside these criteria if, for example,
conditions deteriorate after commencing an MSB but separate QC
procedures result in the removal of poor quality data (see section
\ref{sec_qc}).

Observations comprise four 10\,s Exposures, each offset in a rhombic
dither pattern, which, at the largest extent, measures 6.4\,arcsec
across.  This matches the dither pattern used by the LAS in bands $Y$,
$H$ and $K$ and results in Exposure frames with whole-pixel offsets
which are stacked to make a Stack frame. As in the original UKIDSS
surveys, 10\,s exposures are motivated by reaching a compromise between
several different considerations, such as being short enough to allow
multiple exposures for median filtering and to minimise saturation by
bright sources, but long enough to reduce the contribution of read
noise to photometric errors. 10\,s Exposures also maintain a
relatively low observing overhead.

We note that the $J$ band strategy adopted by the rUHS does not match
the $J$ band strategy used for LAS observations which were
micro-stepped \citep[see][]{dye06}. There are therefore no interleaved
frame types in the WSA for rUHS data and so images have the native
WFCAM pixel scale of 0.4\,arcsec. Since the median PSF FWHM is
0.75\,arcsec in this release, approximately half of all images exceed
the Nyquist sampling limit.  We also note that the rUHS does not
include any bad weather regions which in the LAS were compensated for
by adjusting exposure times.  Following the UKIDSS LAS, GPS and GCS,
$J$ band observations use WFCAM's correlated double sampling read
mode. The procedure starts with a detector reset and then an initial
read. After the desired length of exposure (i.e. 10\,s for rUHS), a
final read is made and the difference between the two reads
constitutes the detected flux, output as counts by the
analogue-to-digital converter.

The 24 rUHS projects are scheduled automatically by the observing
management system at UKIRT. Priorities between projects are set both
dynamically by their airmass and statically using a priority assigned
by the project scientist to ensure an approximately uniform rate of
completion across projects. In addition to the 24 pre-defined
projects, a 25th 'patch-up' project exists with the highest priority.
The purpose of this patch-up project is for the reinstatement of
failed observations as identified by the pipeline
processing. Reinstatement of projects takes an average time of
approximately 2 months from initial acquisition to re-scheduling,
hence the affected Tiles are typically still observable. A whole Tile
is reinstated to ensure a sufficient number of Exposures for the sky
correction frame. This `QC-on-the-fly' is a separate process from the
bulk QC performed prior to data release (see section \ref{sec_qc}).

\subsection{Pipeline data processing}
\label{sec_pipeline}

The data pipeline follows the UKIDSS model where processing is carried
out in two stages using the VISTA/WFCAM data-flow system
\citep{irwin04,hambly08}. In the first stage, discussed in this
section, raw survey data are electronically transferred from UKIRT to
the Cambridge Astronomical Survey Unit (CASU). Here, processing
removes instrumental signatures from the individual Exposures, stacks
Exposures and extracts sources to produce catalogues for each Stack
frame. In the second stage, discussed in section \ref{sec_archive},
catalogue data are integrated into a seamless survey product and
together with the processed image data, are ingested into a database
hosted by the WSA.

\subsubsection{Image processing}

Images are first flattened with the master twilight flatfield which is
updated regularly at approximately monthly intervals. Flattened images
are subsequently sky subtracted using a sky frame computed as a
running average of several median filtered science frames. These
science frames are selected to lie within as narrow a window as
possible around the observation being corrected. Sky frame subtraction
removes residuals due to scattered light and thermal emission mainly
from marks and dirt on WFCAM's field lens.  rUHS MSBs contain eight
Tiles which usually provide ample science frames for median filtering
out sources for the sky subtraction frame. However, in principle, the
pipeline allows these science frames to be selected from any observer
programme so long as they match by filter and exposure time. This
latter requirement ensures that exposure-dependent artifacts are
removed from images as effectively as possible although in a few per
cent of images, sky subtraction residuals are still visible
\citep[see][for further details]{warren07}.

A feature of the detectors in WFCAM is cross-talk which occurs between
the eight parallel channels in each quadrant (see Figure
\ref{wfcam_focal_plane}). This behaviour is similar across all
detectors and produces gradient images in neighbouring channels when
very bright and saturated stars are present. Cross-talk images
typically induce features at $\sim$1 per cent of the differential flux
of the source in adjacent channels, falling to $\sim$0.2 and$\sim$0.05
per cent in successive channels further out. A cross-talk suppression
module removes these artifacts although some residuals may still be
present in the case of extremely saturated sources.

The WFCAM detectors also exhibit the phenomenon of latency, often
referred to as persistence. The effect leaves a residual signal in
exposures taken shortly after previously imaging a very bright,
saturated source. This decays with subsequent exposures but because
the region of the detector remains the same each time, this can
sometimes appear in the sky frame and therefore can manifest itself in
the sky-subtracted science frames. Currently, the pipeline does not
correct for latency.

\subsubsection{Source catalogues}
\label{sec_src_cats}

Source catalogues are generated using a source extraction algorithm
based on the original procedure outlined in \citet{irwin85}. This
identifies sources by searching for groups of four or more
inter-connected pixels which all lie 50 per cent above the local
background level. The background is estimated by bi-linearly
interpolating a grid of sigma-clipped median values that are computed
within $64 \times 64$ pixel blocks over the image \citep[see][for
  further details]{irwin84}. This background estimate is subtracted
prior to source photometry and thus removes slowly varying background
gradients.

A comprehensive range of parameters are determined for each extracted
source.  This includes a suite of fixed aperture magnitudes and, for
extended sources, morphologically-based aperture magnitudes. Apertures
are `soft-edged', meaning that image pixels bisected by the aperture
boundary contribute to the summed flux by the pixel fraction contained
within the aperture. Also included is a morphological classification
which assigns the probability of a source being a star, estimated
using a curve of growth determined from the different fixed aperture
fluxes. Finally, an error flag warns of potential problems in the
photometry due to issues such as bad pixels within the aperture,
channel bias problems (i.e. anomalous offsets between detector
channels), proximity to other sources and whether the source is
saturated (see section \ref{sec_data_access}). Readers are referred to
the schema browser on the WSA for more details and a complete list of
parameters.

\subsubsection{Photometric calibration}
\label{sec_photom_calib}

Photometric calibration is carried out by the CASU data reduction
pipeline. The calibration is based on the magnitudes of 2MASS stars in
the WFCAM photometric system as described in \citet{hodgkin09}. By
matching to stars detected in 2MASS with a positional tolerance of
1\,arcsec, each WFCAM detector frame is calibrated by computing the
median of all per-star zero points (see below for their
definition). This requires transforming the matched stars' magnitudes
from the 2MASS photometric system to the WFCAM system
\citep[see][]{hodgkin09}. The calibration includes a
position-dependent correction applied to all fluxes and magnitudes to
allow for the astrometric distortion which results in a variable
angular pixel scale (see section \ref{sec_astrom}).

The use of WFCAM system magnitudes of 2MASS stars for photometric
calibration in the manner outlined is a departure from more
traditional schemes which are prone to additional systematic errors
whereby standards are observed off-target.  The selection of 2MASS
stars used in the calibration applies an extinction-corrected colour
cut of $0.0 \leq J - K \leq 1.0$ and a signal-to-noise cut of
$>10$. Around 99.9 per cent of detector frames have between 25 and
1000 2MASS stars satisfying these criteria within them. If fewer than
25 2MASS stars are found within a detector, then the colour cut is not
applied.  This process achieves a photometric accuracy of $\sim
2$\,per cent, this figure having been established through analysis of
repeat measurements within UHS data (see section
\ref{sec_data_quality}) and combining this with the 2MASS global
photometric calibration accuracy of $\sim 1$\,per cent \citep{nikolaev00}.

Zero points are defined as the Vega magnitude corresponding to a total
corrected flux of 1 count/s and assume an airmass of unity. These
are stored in the WSA as the keyword {\tt MAGZPT} in the
headers of Flexible Image Transport System (FITS) files and as {\tt
photZPCat} in table {\tt MultiframeDetector}. The uncertainty in
these quantities are {\tt MAGZPTERR} and {\tt photZPCatErr}
respectively and include the 2MASS calibration error, the error
due to uncertainty in the conversion from 2MASS to the WFCAM filter
system and any residual offsets between the detectors \citep[see][for
further details]{hodgkin09}.  The calibrated and corrected magnitude
of a source is therefore related to its observed flux via
\begin{eqnarray}
m = {\tt photZPCat} - 2.5 \log_{10}({\tt aperFlux3}/{\tt expTime}) 
\nonumber \\
- {\tt aperCor3} - {\tt extCorr} - {\tt distCorr}\, , \nonumber
\end{eqnarray}
where {\tt aperFlux3} is the flux in a fixed aperture of radius
1\,arcsec in counts (stored in the WSA source detection table {\tt
  uhsDetection}; see section \ref{sec_data_access}), {\tt expTime} is
the integration time in seconds (stored in {\tt Multiframe}), {\tt
  aperCor3} is the corresponding aperture correction for point sources
to transform to total flux (stored in {\tt MultiframeDetector}), {\tt
  extCorr} = 0.05(({\tt amStart}+{\tt amEnd})/2-1) corrects the
assumption of unit airmass in {\tt photZPCat} (the attributes for
airmass, {\tt amStart} and {\tt amEnd}, being stored in {\tt
  Multiframe}) and {\tt distCorr} is the correction for the image
distortion, computed from astrometry information in the FITS image
datafiles.

\subsubsection{Astrometric calibration}
\label{sec_astrom}

For the astrometric solution, the astrometric distortion introduced by
the optics is first removed with a radially symmetric polynomial model
centred on the WFCAM focal plane \citep[see][]{irwin04}. The
distortion introduced by WFCAM introduces a linear scale change of 0.6
per cent at the edge of the FOV compared to the centre.  In each
detector frame, bright but unsaturated stars are identified and used
to determine the transformation to align with 2MASS which itself uses
the Tycho system \citep{hog00}. For the image data, the world
co-ordinate system is then computed in the ZPN projection
\citep{calabretta02} and the appropriate keywords written to the
multi-extension FITS files.

As with UKIDSS, rUHS astrometry exhibits an expected decrease in
precision with increasing Galactic latitude due to a decrease in the
number of 2MASS stars available for solution (see Section
\ref{sec_data_quality} for more details).

\subsection{Data Archive}
\label{sec_archive}

The processed images and catalogue products generated by CASU in the
first stage of the pipeline are delivered to the Wide Field Astronomy
Unit (WFAU) in Edinburgh. At this point, every detector frame has an
individual associated catalogue. In the second stage of processing at
WFAU, these catalogues are merged into a seamless catalogue
product. This procedure merges duplicated sources extracted from
overlapping detector frame regions, and, although not relevant for the
present $J$-only release, also cross-matches sources detected in
different wavebands. This preparation of catalogue data includes a
thorough quality control phase which we describe in detail in Section
\ref{sec_data_props}.  

Processed images and merged catalogue products
are then ingested into a relational database hosted by the WSA at {\tt
  http://wsa.roe.ac.uk}.  The reader is referred to \citet{hambly08}
for more details on the second stage of the pipeline regarding
catalogue preparation as well as the design and structure of the
database. In addition, the WSA web pages provide comprehensive
information on how data are stored and accessed. We provide an
overview of how to obtain basic data products in section 
\ref{sec_data_access}.

\section{Data release products}
\label{sec_products}

In this section, we specify the data product types made available in
this release and how to access them. Section \ref{sec_data_props}
gives details of the characteristics of the data being released.

UKIDSS survey data are already available via the WSA. The rUHS
$J$-band data being made available in this release form a distinct,
separately queryable database in the WSA called {\tt UHSDR1}. By early
2018, existing UKIDSS data will have been merged with rUHS data but at
the present time, users must execute separate queries within rUHS or
UKIDSS depending on the co-ordinates being targeted.

\subsection{Image data}

Image pixel data can be obtained from the WSA in several
ways. Menu-driven searches allow users to list Exposure and Stack
multiframes which coincide with a given position. These can be
downloaded as multi-extension FITS data files with the option of being
compressed or uncompressed. Similarly, an image cut-out service
provides pixel data of a user-defined geometry around an input
co-ordinate or an uploaded list of co-ordinates up to a maximum size
of $15 \times 15$ arcmin. Note that unlike some of the $J$-band
observations present in the WSA acquired under the UKIDSS programme,
there are no interleaved image products and therefore there are no
{\it leavestack} datatypes for rUHS in the WSA.

Image data may also be obtained using free-form SQL queries.
Primarily, this is accomplished by the attribute {\tt fileName}
in the table {\tt Multiframe} which directly links to
the multi-extension FITS file stored on the server (see
section \ref{sec_data_props}).

The release includes $\sim 65$,200 Stack and $\sim 261$,000 Exposure
multiframes which have passed QC criteria. This corresponds to an
areal coverage of $\sim 12$,450deg$^2$ by Stack frames
once overlaps are taken into account, $\sim 98$\,per cent of the
nominal rUHS area. As we discuss further in section
\ref{sec_dr1_specifics}, this data release also includes data
which have failed the QC criteria. This extended dataset
covers 99.9\,per cent of the rUHS area.

\subsection{Source catalogues}

Source catalogues can be obtained with a number of different methods.
As with access to image pixel data, menu-driven web interfaces allow
sources to be searched for within a user input co-ordinate.
The web interface also allows users to upload an external source
catalogue for cross-matching against UHS sources. 

Free-form SQL queries enable generation of user-specific source
catalogues. For example, sources within a given range of RA, Dec,
magnitude, size, ellipticity or more complicated combinations of
source properties can be selected.

\subsection{Access to data products}
\label{sec_data_access}

The relational data base at the heart of the WSA is interfaced via
structured query language (SQL). Users can query the data base using
direct free-form SQL or menu-driven searches. The menu-driven searches
allow users to download the processed FITS image pixel data (for both
Exposures and Stacks) and source catalogues by matching to uploaded
co-ordinate data (see section \ref{sec_products}).  Free-form SQL
queries enable a much greater degree of flexibility and some useful
SQL examples for users are given in \citet{dye06} and
\citet{hambly08}.

The WSA database essentially contains a collection of
tables. Information on Exposure and Stack multiframes (referred to in
the WSA as `Normal' and `Stack' multiframes respectively), both of which
comprise an image for each of the four WFCAM detectors as discussed in
section \ref{sec_geom}), is stored in a table called {\tt
  Multiframe}. This table contains attributes that can be queried,
such as the frame type ({\tt frameType}, which for the rUHS will
usually be 'normal' or 'stack'), the base co-ordinates of the
observation ({\tt raBase, DecBase}), the filter used ({\tt filterID}),
the airmass ({\tt amStart} and {\tt amEnd}) and the observation date
({\tt dateObs}). {\tt Multiframe} also contains the attribute ({\tt
  fileName}) which provides a web link to the processed pixel data
stored as a FITS file for direct download. Every Normal and Stack
multiframe has the unique attribute {\tt multiframeID} which is common
to many other tables in the WSA to allow cross-linking.

Whilst {\tt Multiframe} contains global information common to each of
the four detector frames that constitute the Normal or Stack frame,
the table {\tt MultiframeDetector} gives information specific to each
detector frame. {\tt MultiframeDetector} is linked to the {\tt
  Multiframe} table via the attribute {\tt multiframeID}.  Individual
detectors are referenced within {\tt MultiframeDetector} via the {\tt
  extNum} attribute (which labels detectors 1,2,3 and 4 as 2,3,4 and 5
respectively, in alignment with the FITS header labelling).  The table
holds data on the photometric calibration (see section
\ref{sec_photom_calib}) and image quality such as the FWHM
of the PSF (given as the attribute {\tt seeing}),
stellar ellipticity ({\tt avStellarEll}) and the median sky brightness
({\tt skyLevel}).

Source catalogue data are stored in one of two main tables. The table
{\tt uhsDetection} stores all sources detected in all Stack frames
included in the standard release products (i.e. not sources detected
in duplicated frames).  Although not relevant for this data release, a
source detected in multiple passbands will appear multiple times in
{\tt uhsDetection}. In addition, this table will contain more than one
entry for the same source if it lies within a region where Stack
frames overlap. In contrast, the table {\tt uhsSource} is constructed
by cross-matching between passbands. In this way, a source not located
within an overlapping Stack frame region will contain only one entry
in {\tt uhsSource} with multi-band photometry. In the present release,
this table holds only $J$-band detections.  Sources common to two or
more overlapping Stack frame regions are matched using a positional
tolerance of 2\,arcsec and then flagged in {\tt uhsSource} using the
{\tt priOrSec} flag (see the WSA schema browser for more
details). This flagging designates the primary source as that which,
firstly, has the least severe errors (stored for each source in the
{\tt ppErrBits} flag -- see below) and, secondly, is furthest from the
detector edge. The table {\tt uhsSource} contains calibrated
magnitudes whereas {\tt uhsDetection} holds both calibrated magnitudes
and fluxes.  Catalogues use equatorial (J2000.0 equinox), Galactic and
SDSS $(\lambda,\eta)$ co-ordinates \citep{stoughton02}.

Sources contained in both {\tt uhsSource} and {\tt uhsDetection} are
assigned an error flag to indicate whether there are likely to be
problems with the photometry or morphological parameters.  In {\tt
  uhsDetection}, this flag is the attribute {\tt ppErrBits} whereas in
{\tt uhsSource}, the flag is given a prefix corresponding to the
filter it was detected in, for example, in this data release, only the
flag {\tt jppErrBits} is populated. The flag is used to warn that
there are bad or saturated pixels in the flux aperture, that the
source lies in a region with poor flatfielding or within a dither
offset of the stacked frame boundary, or that the source may be
affected by cross-talk.  The WSA schema browser gives further details.

Within the database, every source has an associated neighbour table to
allow more sophisticated matching algorithms, for example, to identify
moving sources where multi-epoch observations are available or to
determine clustering statistics. This is implemented by the {\tt
  uhsSourceNeighbours} table which has been generated by matching {\tt
  uhsSource} to itself and recording all neighbours within 10\,arcsec
for every source. Sources can also be matched to external source
catalogues. A selection of tables has been constructed in the WSA to
allow such cross-matching. For example, {\tt uhsSourceXDR13PhotoObj}
and {\tt uhsSourceXwise\_allskysc} match to SDSS DR13
\citep{albareti17} and WISE \citep{wright10} respectively.

\subsection{Data release specifics}
\label{sec_dr1_specifics}

Included in this release are additional image and source catalogue
data which have not been used to generate the standard release
products ({\tt uhsSource and uhsDetection}).  These `extended release'
data products fall into one of two categories.  The first is data
which have not met QC criteria, mainly due to a poor PSF FWHM and/or poor
depth (see section \ref{sec_qc} for more details). The second is data
which may or may not have passed the QC criteria but have been
excluded from the standard release products because they are repeat
observations and their inclusion in the archive ingest process causes
significant delay due to the triggering of deep stacking, source
extraction and ingestion.  (A later data release will include deep
stacking of these repeat areas.)  This second category is primarily
due to repeat observations which have been carried out when an MSB has
been stopped and restarted by a telescope operator but also a
significant fraction ($\sim 30$\,per cent) were to rectify a problem
with incomplete Tiles (see below). In cases where multiple repeat
observations at the same base position pass all QC criteria, the
deepest Stack and its associated Normal frames were used to generate
the standard release products.

\begin{figure*}
\epsfxsize=160mm
{\hfill
\epsfbox{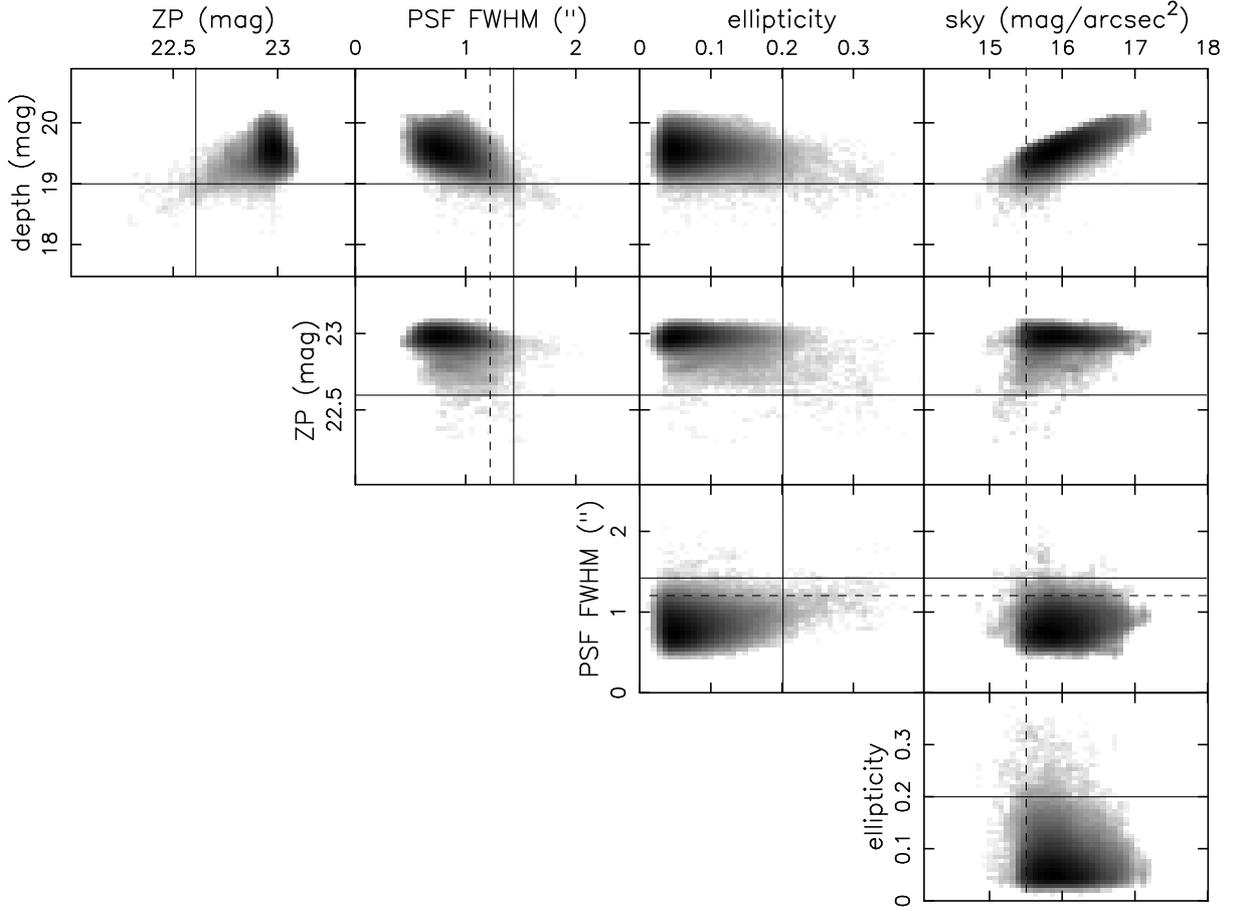}
\hfill}
\caption{Correlations between data quality measurements for each
  detector frame in this release. Grey-shading shows the logarithm of
  the number density of detector frames at a given point. Quantities
  are $J$-band depth (Vega magnitudes), PSF FWHM (arcsec), stellar
  ellipticity, zero point magnitude and sky brightness ($J$-band
  mag/arcsec$^2$). See section \ref{sec_qc} for definitions of these
  quantities. Solid lines indicate the cut applied in the QC process
  whereas dashed lines show the site quality constraints imposed for
  observations.}
\label{quality_corr}
\end{figure*}

The images included in these additional deprecated data are listed
within the table {\tt Multiframe} alongside the standard release image
data.  All images and catalogue products can be retrieved
from the archive using the 'Archive Listing' form.  However, other
menu-driven WSA searches revert to using the standard products {\tt
uhsSource uhsDetection} which excludes these additional frames. Users
must therefore either deselect the 'purge deprecated frames' option
when using menu-driven searches or query the archive with free-form
SQL to access them. The attribute {\tt deprecated} in {\tt Multiframe}
and {\tt MultiframeDetector} is assigned a value of zero if the frame
is included in the standard release or, if it has been excluded, a
code greater than zero indicating why it has not be included (see the
schema browser on the WSA web pages for more details).

To incorporate the sources detected in these additional frames, two
extra tables, {\tt uhsDetectionAll} and {\tt uhsSourceAll}, have been
created. The table {\tt uhsDetectionAll} contains all sources already
held in {\tt uhsDetection} plus all additional sources detected
in the extra deprecated images regardless of whether they have failed
QC criteria or not.  When querying {\tt uhsDetectionAll}, the
attributes {\tt deprecated} and {\tt ppErrBits} indicate whether a
source originates from an image included in the standard release or not
and why. The table {\tt uhsSourceAll} includes all sources from {\tt
uhsSource} plus only those additional sources which originate from
frames that did not meet QC criteria. Sources extracted from repeated
frames which pass QC are not included in the merging process and thus
are not included in {\tt uhsSourceAll}. Both {\tt uhsDetectionAll} and
{\tt uhsSourceAll} can not be queried with menu-driven searches but
must be accessed using free-form SQL only.

An issue that arose in the early stages of rUHS operations was that a
software bug in the data acquisition system introduced after an
upgrade at UKIRT resulted in the formation of incomplete Tiles. The
effect was to observe only three out of the four required base
positions with one of these randomly being imaged twice. This
affected $\sim 650$ Tiles randomly. The bug was corrected on September
14th 2012 and the problem eliminated thereafter. The affected Tiles
were re-queued and subsequently re-observed in full. All four base
positions of affected Tiles were observed rather than solely the
missing base position. This was to ensure a sufficient number of
Exposures observed under similar conditions and close together in time
for the creation of a good quality sky frame.

Although these repeat frames have not been deep stacked, these
additional data are included in the release and in principle offer an
increase in depth of $\sim 0.4$\,mag over a non-contiguous area of
$\sim 390$\,deg$^2$ compared to the nominal depth. Since the
incomplete Tile problem also duplicated one base position at random,
within this area of increased depth, approximately 130\,deg$^2$ is
deeper than nominal by $\sim 0.6$\,mag. In addition to the increase in
depth, these repeat observations allow multi-epoch study with a
baseline of up to four months. 

\begin{figure}
\epsfxsize=75mm
{\hfill
\epsfbox{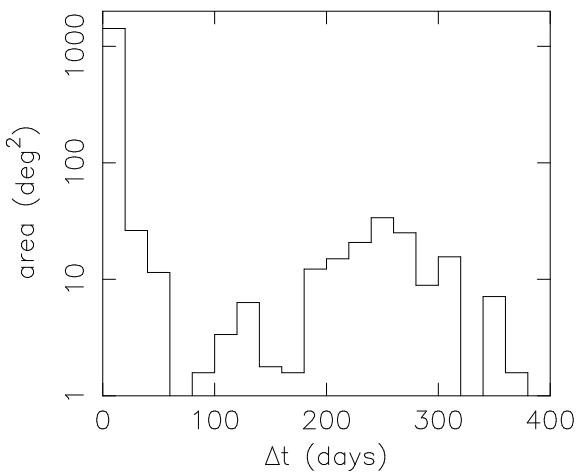}
\hfill}
\caption{Areal coverage of repeats in this release binned by the time
  elapsed between the first and last observation within a group of
  repeat observations. This amounts to a total area of $\sim
  1630$\,deg$^2$ and has been computed for all data in the release,
  including data that have not passed QC but are included in the
  extended release. Within this, an area of approximately
  1410\,deg$^2$ has repeat imaging where at least two of a group of
  repeats pass QC.}
\label{epoch_distn}
\end{figure}

Figure \ref{epoch_distn} shows the areal coverage of all detector
frames repeated at least once, as a function of the maximum separation
in time between repeats.  The figure applies to all repeated data,
including re-observations for the incomplete Tile problem and repeats
resulting from re-queueing of aborted but partially completed
MSBs. These data amount to a total of area of $\sim 1630$\,deg$^2$
with at least two stacked detector frames but including data that have
failed QC. If this is limited to data where there are at least two
repeated detector frame stacks that have both passed QC, then this
area falls to $\sim 1410$\,deg$^2$.

\section{Data characteristics}
\label{sec_data_props}

In this section, we outline the QC criteria applied to the image and
source catalogue products released. Section \ref{sec_data_quality}
summarises characteristics of the data.

\subsection{QC criteria}
\label{sec_qc}

All rUHS data made available in this release have been quality-checked
using a QC process involving three stages.  The first and most
fundamental stage is an automatic data integrity check which
deprecates corrupt data, including any empty detector images or pixel
data with damaged metadata.

In the second stage, all images are manually eye-balled. This process
identifies any poor-quality images which are not picked up by the
quantitative criteria described below. For example, trailed images
often give rise to acceptable measurements of PSF FWHM and ellipticity
since only a small fraction of the source's flux is typically trailed.
Similarly, extreme variation in channel bias or poor sky subtraction
is commonly not manifested in quantities such as frame depth or sky
brightness measurements. Other problems that eyeballing identifies are
poor flatfielding, poor cross-talk correction and 'disasters' which is
a catch-all category for catastrophic failures due to unknown
reasons. Features that are ignored in the eyeballing process include
satellite trails, diffraction spikes from bright sources, saturated
sources, detector latency and small gradients in the background.  This
latter feature can be ignored since the source extraction process
removes local background gradients (see Section \ref{sec_src_cats}).

For this release, eyeballing was shared between a team of nine. Each
team member simultaneously viewed all four detector images belonging
to a given Stack at a time, for a subset of all Stacks. In this
process, poor quality multiframes with problems discussed previously
were flagged.  Consistency within the team was verified by assigning
each member a common but randomly selected fraction of the data.  To
further improve consistency, all flagged frames were then re-eyeballed
by the project scientist to un-flag any frames deemed acceptable and
then apply a code to the unacceptable frames to indicate the nature of
the failure. In this way, the eyeballing procedure deprecates entire
multiframes, not just detector frames. The attribute {\tt deprecated}
in table {\tt Multiframe} in the WSA is given the value of the
eyeballing failure code which ranges from 60 to 70 inclusive.
Similarly, when set, bit number 30 of the flags {\tt ppErrBits} and
{\tt jppErrBits} in tables {\tt uhsDetectionAll} and {\tt
  uhsSourceAll} indicates that a source has been extracted from an
image that has not passed eyeballing checks. We note that only data
designated as being of type 'disaster' or 'empty frame' are completely
removed from the extended data products.

In the third stage of QC, data are subjected to a series of
quantitative thresholds. These result in data being rejected at the
detector frame level rather than the multiframe level as with
eyeballing.  For this release, thresholds were applied to PSF FWHM,
stellar ellipticity, zero point and frame depth as follows:
\begin{itemize}

\item[] {\it PSF FWHM} -- The rejection threshold for PSF FWHM is set
  at 1.4\,arcsec as measured from point sources averaged over the
  entire detector frame. This is greater than the site quality constraint
  for the PSF FWHM of 1.2\,arcsec to allow for slight broadening of the PSF
  by sub-optimal guiding and also to allow for changes in seeing whilst an
  MSB is being executed by UKIRT.

\item[] {\it Depth} -- Depth is defined as the total magnitude of a
  point source for which the integrated flux within a 2\,arcsec
  diameter aperture is detected at the $5\sigma$ confidence level.
  The threshold applied in this release is at a $J$-band magnitude of
  19.0\,mag corresponding to 0.6\,mag shallower than the median value.

\item[] {\it Stellar ellipticity} -- Detector frames with a mean stellar
  ellipticity ({\tt avStellarEll} in the table {\tt
    MultiframeDetector}) of $>0.2$ averaged over the detector frame are
  rejected. Stellar ellipticity is defined as $1-b/a$ where $a$ and
  $b$ are semi-major and semi-minor axis length respectively. This
  threshold matches that applied in the UKIDSS LAS. Although the value
  of 0.2 may be regarded as being quite high, as we show in the next
  subsection, frame depth is not significantly compromised when stellar
  ellipticity approaches this limit. Values larger than the threshold
  usually indicate that the data suffer from additional issues and
  these are typically picked up by the other cuts applied.

\item[] {\it Zero point} -- The threshold applied to zero point ({\tt
  photZPCat} in table {\tt MultiframeDetector}) is set 0.4\,mag
  brighter than modal. This is a relatively loose threshold and for
  this release corresponds to the limit 22.6\,mag.  This rejects
  detector frames with outlying zero points whilst relying on the
  depth threshold to provide the primary criterion.  The philosophy
  here is not to cut too severely on a parameter when the depth has
  not been compromised.

\end{itemize}

\subsection{Data quality}
\label{sec_data_quality}

\begin{figure}
\epsfxsize=84mm
{\hfill
\epsfbox{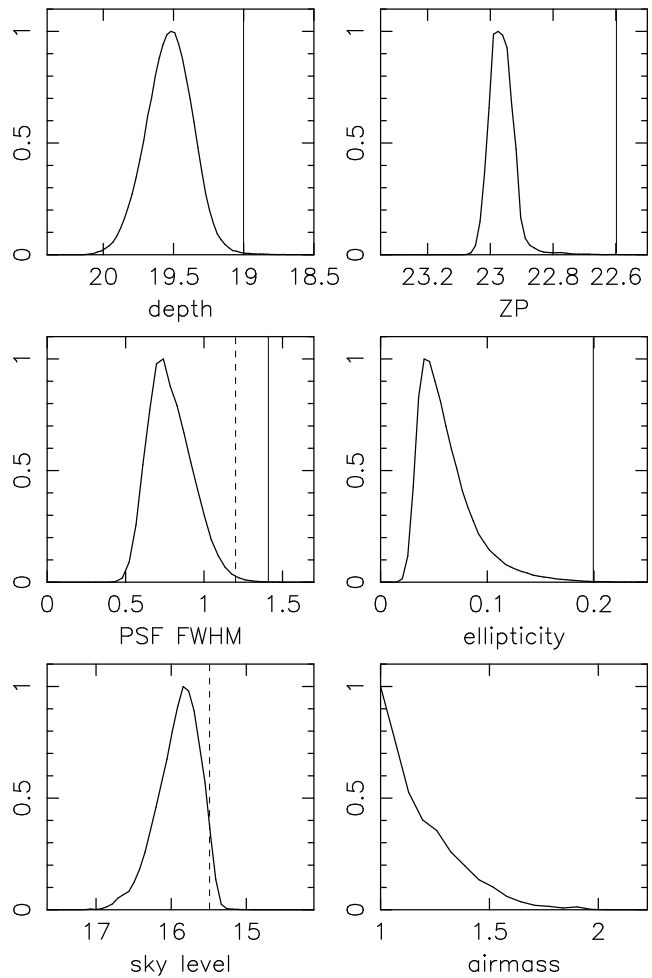}
\hfill}
\caption{Distributions of data quality measurements for all detector
  frames in this release. Quantities are $J$ band depth (Vega
  magnitudes), PSF FWHM (arcsec), stellar ellipticity, zero point
  magnitude, sky brightness ($J$ band mag/arcsec$^2$) and airmass. See
  section \ref{sec_qc} for definitions of these quantities. Solid
  lines indicate the cut applied in the QC process whereas dashed
  lines show the site quality constraints imposed for observations.}
\label{quality_dist}
\end{figure}

Figure \ref{quality_corr} shows the correlation between depth, PSF
FWHM, zero-point, stellar ellipticity and sky brightness for all
305,280 detector frames being released. The plots indicate the
thresholds applied at the detector frame level during the QC process
and the site quality constraints set for observations (also shown more
quantitatively in Figure \ref{quality_dist}). As the figure shows, up
to the ellipticity cut at 0.2, depth is distributed quite
symmetrically about the median value of 19.6\,mag, demonstrating that
depth is not compromised by minor tracking inaccuracies. In addition,
ellipticity only weakly affects image PSF FWHM. However, as expected,
depth is strongly correlated with PSF FWHM and sky brightness.

Figure \ref{quality_dist} shows distributions of the quality
measurements featured in Figure \ref{quality_corr} along with the
thresholds applied in the QC process. The figure additionally shows
the distribution of airmass of detector frames. The thresholds reject
0.8\,per cent by PSF FWHM, 1.3\,per cent by depth and 0.4\,per cent by
zero point. The rate of detector frames rejected by the eyeballing
process is 1.7\,per cent (with $\sim 80$\,per cent of this being
attributed to guiding failures). Combining all of these (largely
mutually inclusive) rejection rates gives an overall rate of 2.3\,per
cent of data having not met quality requirements. To reiterate, apart
from disasters and corrupt data which constitute a negligible
fraction, these rejected data are still available to the user via the
tables {\tt Multiframe}, {\tt uhsDetectionAll} and {\tt uhsSourceAll}
(see section \ref{sec_dr1_specifics}).

\begin{figure}
\epsfxsize=84mm
{\hfill
\epsfbox{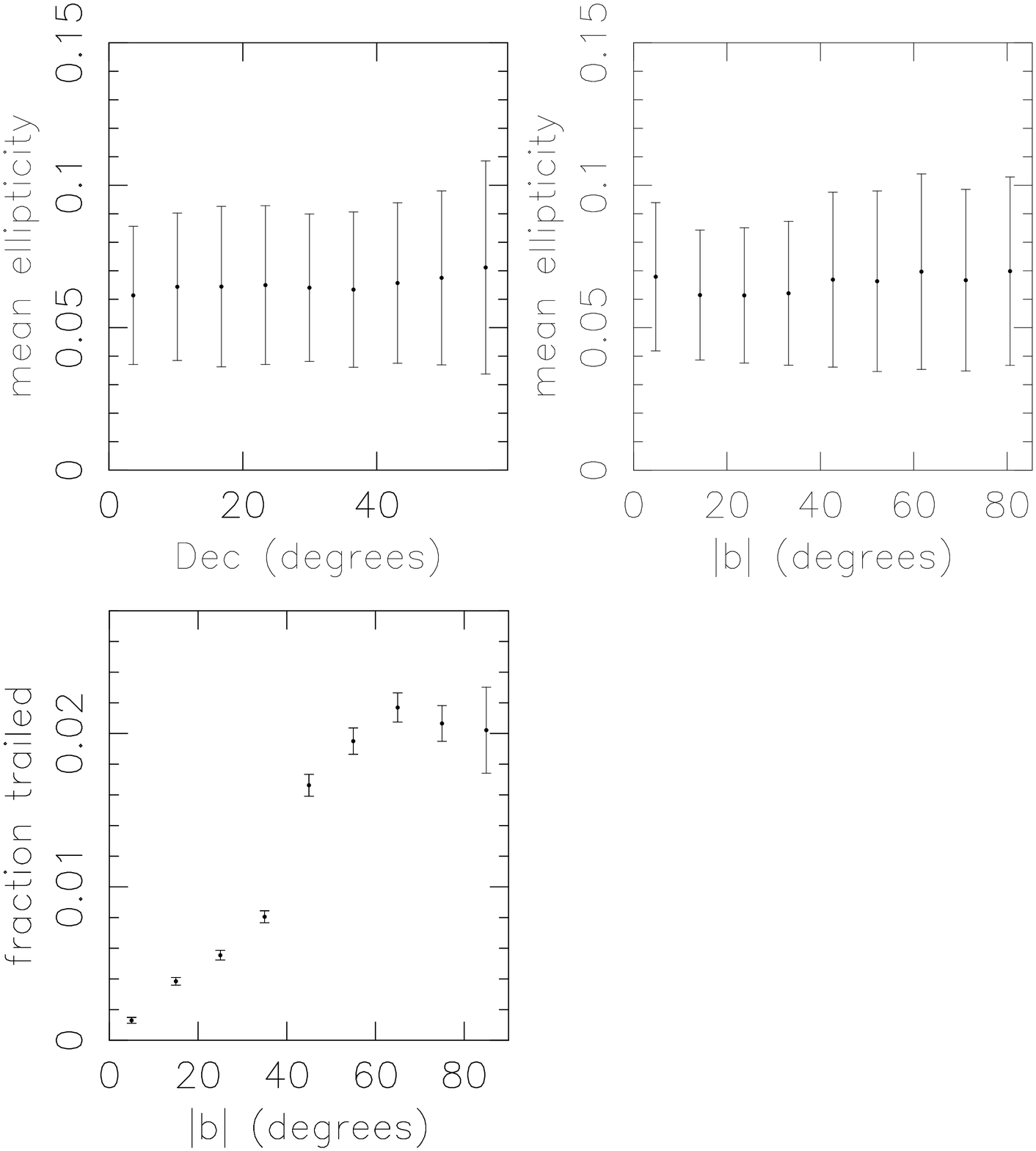}
\hfill}
\caption{{\em Top left:} Variation of stellar ellipticity with
  declination. Error bars show the standard deviation of ellipticity
  in each declination bin. The plot shows negligible degradation in
  image quality at high declination, demonstrating that UKIRT's
  secondary mirror compensates for instrument flexure well (see main
  text for more details). {\em Top right:} Variation of stellar
  ellipticity with galactic latitude. Again, negligible variation
  shows that even at high galactic latitudes (and thus low density of
  guide stars), guiding quality is maintained. {\em Bottom left:}
  Fraction of multiframes that have failed QC by being trailed as a
  function of galactic latitude. Error bars show Poisson noise.}
\label{quality_ellip}
\end{figure}

Some of the earlier UKIDSS data releases reported a positive
correlation between stellar ellipticity and declination. This was
attributed to instrument flexure at high declination. The remedy
implemented for later releases was to compensate for flexure using
UKIRT's secondary mirror. Figure \ref{quality_ellip} shows the
variation of mean stellar ellipticity with declination for data in
this release. Whilst there is a slight elevation in ellipticity with
increasing declination, this is at a significantly lower level than
that reported by UKIDSS and is well within the scatter of
ellipticities measured.

Figure \ref{quality_ellip} also shows that there is negligible
variation in mean stellar ellipticity with galactic latitude. At
higher galactic latitudes, the stellar density falls dramatically,
limiting the choice of stars with suitable signal-to-noise for
guiding. This lack of variation therefore shows that the paucity of
guide stars at higher galactic latitudes has not noticeably affected the
guide quality of QC-passed data. However, the same is not true of the
failure rate due to trailed images. The bottom-left panel of Figure
\ref{quality_ellip} shows the fraction of multiframes that have failed
QC by being trailed as a function of their galactic latitude. Here,
there is a dramatic increase from almost no frames being trailed at
low latitudes to approximately 2\,per cent of frames being trailed at
latitudes $>60^\circ$.  This trend is not seen with declination and
hence cannot be attributed to instrument flexure. Instead, the likely
dominant cause of these guiding failures is contamination of the
selection of guide stars used for rUHS by compact galaxies (see Section
\ref{sec_geom}) and a stronger occurrence of fainter guide stars at high
galactic latitudes.

\begin{figure*}
\epsfxsize=160mm
{\hfill
\epsfbox{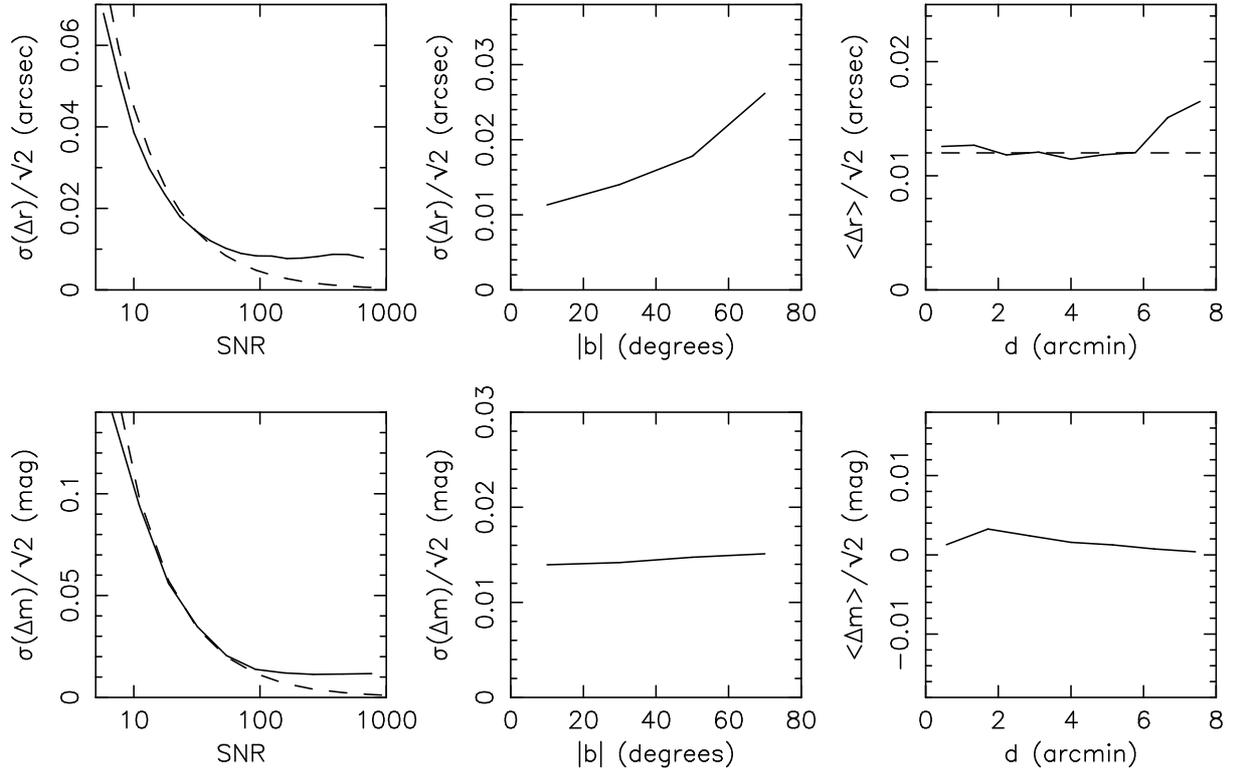}
\hfill}
\caption{Internal consistency of astrometry and photometry determined
  from repeat observations. The top row shows how the standard
  deviation (left and middle plots) and mean (right plot) of the
  separation, $\Delta r$, of unsaturated stars paired between
  duplicated frames varies as a function of their SNR, galactic
  latitude and distance, $d$, from the detector frame
  centre. Similarly, the bottom row shows the variation in standard
  deviation (left and middle plots) and mean (right plot) of the
  magnitude difference, $\Delta m$, of unsaturated stars paired
  between duplicated frames.  The dashed line in the top-left plot
  shows the expected variation in positional precision with SNR as
  given by \citet{ivison07} for a PSF FWHM of 0.75\,arcsec. The dashed
  line in the bottom-left plot shows the theoretical line
  $\sigma_m=2.5/({\rm SNR} \times \ln10)$ for a 1$\sigma$ error in
  magnitude, $\sigma_m$. The dashed line in the top-right plot shows
  the expected mean offset for the measured astrometric error of
  0.01\,arcsec at SNR=100. For the four plots in the middle and
  right-hand columns, stars were selected with $99.5 < SNR < 100.5$.}
\label{internal_consistency}
\end{figure*}

\subsubsection{Astrometric Precision}

The top row of Figure \ref{internal_consistency} gives various
measures of the astrometric precision of rUHS catalogue data by
analysing sources selected from pairs of duplicated detector frames,
where both frames in each pair have passed QC criteria (i.e. these
are largely the frames which were repeated to patch up incomplete
Tiles -- see section \ref{sec_dr1_specifics}). The plot in the top
left of the figure shows how the standard deviation of the separation
between matched pairs of unsaturated stars varies with signal-to-noise
ratio (SNR).  Dividing this by $\sqrt{2}$ gives an estimate of the
radius of the circle within which the source centroid can be located
with a $1\sigma$ confidence (and hence, dividing by a further factor
of $\sqrt{2}$ gives an estimate of the 1$\sigma$ uncertainty in
declination or right ascension). The dashed line is the theoretical
expectation of the radius of this error circle taken from
\citet{ivison07} for the median survey PSF FWHM of 0.75\,arcsec.  This
line agrees well with that measured up until a SNR of approximately
100 where a limit in the astrometric precision of 0.01\,arcsec is
reached due to the precision with which centroids are determined
in the source extraction process.

The middle plot in the top row of Figure \ref{internal_consistency}
shows how astrometric precision varies with galactic latitude for
unsaturated stars selected by $99.5 < SNR < 100.5$. At low latitudes,
the precision is close to the limit of 0.01\,arcsec, but this
increases monotonically to $\sim 0.03$\,arcsec by a galactic latitude
of $b=80^\circ$ where there are fewer 2MASS stars with which to
determine the astrometric solution. Finally, the plot on the
right-hand side of Figure \ref{internal_consistency} shows how the
mean positional offset between pairs of unsaturated stars located in
duplicated frames selected by $99.5 < SNR < 100.5$ varies as a
function of distance from the detector frame centre.  The dashed line
shows the expected mean offset for the measured astrometric error of
0.01\,arcsec for a SNR of 100 taken from the first plot.  There is
good agreement between the expected and measured mean offset within a
radius of 6\,arcmin from the centre of the detector frame ($\sim 60$
per cent of the detector frame area) beyond which, the measured mean
offset rises by $\sim 30$ per cent. Nevertheless, this is extremely
small compared to the global astrometric precision of 2MASS of $\sim
0.2$\,arcsec \citep{cutri00}, demonstrating that systematic errors due
to detector edge effects are negligible.

\subsubsection{Photometric Precision}

Photometric precision was determined with the same catalogue data used
for the astrometric analysis, i.e., unsaturated stars selected from
duplicated detector frames. The bottom row of Figure
\ref{internal_consistency} shows the results of this analysis. In the
bottom-left plot of the figure, the standard deviation of the
difference in $J$-band magnitude between matched pairs of stars is
seen to vary with SNR in a manner that closely matches the variation
expected from simple error propagation (such that the 1$\sigma$ error
in magnitude varies with SNR as $2.5/({\rm SNR} \times \ln10)$) up to
a SNR of 100. At higher SNRs, the photometric precision reaches an
asymptotic value of $\simeq 0.01$\,mag. Combining this limiting
internal photometric precision with the global 2MASS precision of
$\sim 1$ per cent \citep{nikolaev00} gives an absolute photometric
precision for the $J$-band data in this release of better than 2\,per
cent.

The middle plot in the bottom row of Figure \ref{internal_consistency}
shows that photometric precision is almost consistent across all
galactic latitudes, rising only $\sim 15$\,per cent from low to high
latitudes. Similarly, the plot in the bottom-right of Figure
\ref{internal_consistency} shows negligible systematics in photometric
precision due to detector edge effects.

\begin{figure*}
\epsfxsize=180mm
{\hfill
\epsfbox{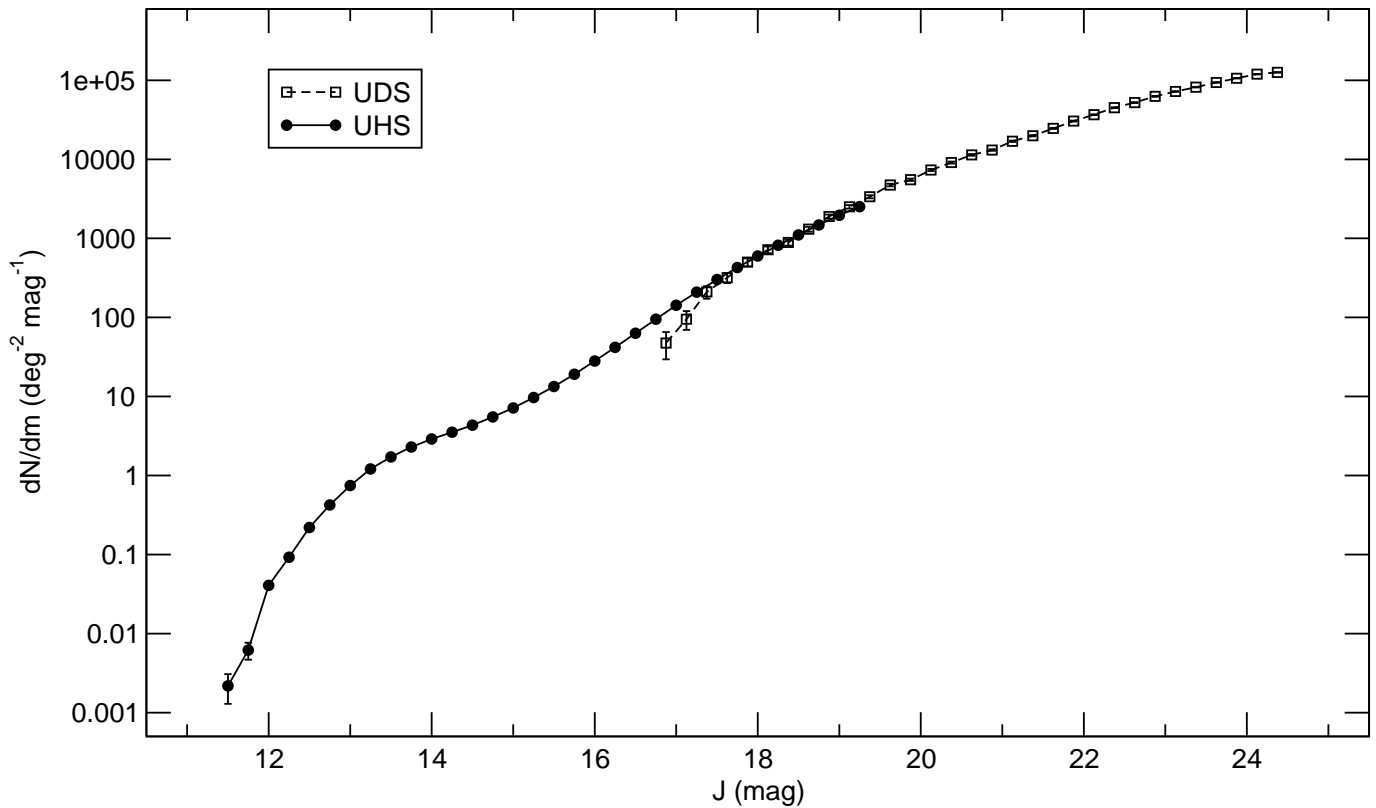}
\hfill}
\caption{Comparison of $J$ band galaxy number counts from the rUHS data
  in this release and the UKIDSS UDS DR11 (Almaini et al., in
  prep). To determine rUHS galaxy counts, unsaturated sources were
  selected from a region defined by $|b|>30^\circ$ and classified by
  the WSA attribute {\tt jClass} as being non-stellar. For UDS,
  galaxies were distinguished from stars using optical imaging in a
  $\sim 0.6$\,deg$^2$ region of the $\sim 0.8$\,deg$^2$ UDS field.
  Error bars account only for Poisson noise.}
\label{number_counts}
\end{figure*}

\subsubsection{$J$-band Number Counts}

As a final illustration of the extent of the data included in this
release, Figure \ref{number_counts} shows the number counts of
galaxies detected in the rUHS in comparison with those determined from
the UKIDSS Ultra Deep Survey (UDS; Almaini et al. in prep).  To
measure the rUHS galaxy number counts, sources were selected from the
WSA table {\tt uhsSource} in a region defined by $|b|>30^\circ$, being
classified by the WSA attribute {\tt jClass} as being non-stellar and
being unsaturated according to the attribute {\tt jppErrBits}. It is
likely that there is a significant contamination of the resulting rUHS
number counts from stars. Quantification of stellar contamination is
beyond the scope of such a simple demonstration of dynamic range and
is therefore left for future work. For the UDS counts, stars were
removed using colour cuts computed using an overlapping $\sim
0.6$\,deg$^2$ optically imaged region of the full $\sim 0.8$\,deg$^2$
field (details to appear in Almaini et al., in prep). The error bars
plotted allow only for Poisson noise and do not account for variation
in PSF FWHM which will affect the 2\,arcsec diameter aperture
magnitudes used (although the median rUHS PSF FWHM is within 5\,per
cent of that of the UDS), the inconsistent star-galaxy separation, nor
cosmic variance. The UDS galaxy counts at the bright become discrepant
with the rUHS counts since bright galaxies in the UDS catalogue have
been removed. Despite the ramifications of these caveats, agreement is
very good in the overlap region.

\section{Summary}
\label{sec_summary}

This paper has defined the UKIRT Hemisphere Survey (UHS) and the
release of $\sim 12,700$\,deg$^2$ of new $J$-band survey data. In
combination with existing UKIDSS data, this completes the full UHS
$J$-band coverage of $\sim 17,900$\,deg$^2$, although $\sim 250$\,deg$^2$
has failed quality control and will be re-observed within one year
of this release.

The newly released data take the form of $J$ band imaging and source
catalogue products.  A total of approximately 500 million unique
sources have been detected, reaching a median 5$\sigma$ point source
sensitivity of 19.6\,mag (Vega) measured within a 2\,arcsec diameter
aperture. The median PSF FWHM of the new image data products is
0.75\,arcsec. In addition, a small, non-contiguous area of $\sim
1630$\,deg$^2$ has been repeated, offering an increase in depth by
$\sim 0.4$\,mag over this area. These additional frames have not been
deep-stacked in this release but will be provided in a future release.

The data were released via the WSA at {\tt wsa.roe.ac.uk} on August
1st 2017 to UK, University of Hawaii and University of Arizona
astronomers.  A world-wide release is scheduled for August 1st 2018.
In this present release, the new data are archived in a distinct
database stored at the WSA. In a future release, planned for early
2018, the data will be merged with existing UKIDSS LAS, GPS and GPS
data to provide a single, seamless database spanning a contiguous area
in the Northern hemisphere up to a declination of 60$^\circ$.

At the time of writing, UHS operations are continuing with
re-observation of the $J$ band Tiles that failed the quality
criteria. Running in parallel with this, $K$ band imaging started in
late-July 2017 at which time survey management was transferred to the
United States Naval Observatory. UKIRT is now commencing a more
survey-oriented mode of operation and so it is anticipated that the
full $K$-band sweep will be completed within a two-year time frame.

\section*{Acknowledgements}

SD acknowledges support by the UK STFC's Ernest Rutherford Fellowship
scheme. ACE acknowledges support from STFC grant ST/P00541/1.  The UHS
data in this release have been acquired under an evolving
collaboration which has involved the following international partners:
STFC, The University of Hawaii, The University of Arizona, Lockheed
Martin and NASA.  The authors wish to recognise and acknowledge the
very significant cultural role and reverence that the summit of Mauna
Kea has always had within the indigenous Hawaiian community. We are
extremely grateful to have the opportunity to conduct observations
from this mountain.

\label{lastpage}

\end{document}